\documentstyle[editedbook,epsfig,psfig,epsf]{mq}

\def\etal{{\em et al.} }

\def\msun{$M_{\odot}$ }

\def\ang{\AA }  
\def\cm2{cm$^2$ }
\def\se1{s$^{-1}$ }

\def\degree{$^{\circ}$}

\newcommand{\ha}{H$\alpha$ }
\newcommand{\hb}{H$\beta$ }

\begin{opening}
\title{Rarefied Gaseous Disk Around Black Hole in the System of V4641 Sgr}
\author{V.P. Goranskij$^1$, E.A. Barsukova$^2$, A.N. Burenkov$^2$, \& D.N. Monin$^2$}
\institute{$^1$ Sternberg Astronomical Institute, Moscow University, Moscow,
119992, Russia.\\
$^2$ Special Astrophysical Observatory, Nizhny Arkhyz, Karachai-Cherkesia,
369167, Russia.
}
\end{opening}

\runningtitle{Gaseous Disk in V4641 Sgr}
\runningauthor{Goranskij, Barsukova, Burenkov \& Monin}
\begin{document}
\vspace{-0.5cm}

\begin{abstract}
{\small
The results of photometric CCD monitoring and spectral observations
of the black hole binary and microquasar V4641 Sgr in the quiet
state are presented. The ellipsoidal light curve with large amplitude
of 0$^m$.36 in $R$ band suggests the influence of a massive object orbiting
around a normal B9 star. In the spectra taken with the 6-m telescope
one hour before black hole inferior conjunction, an absorption component
in the red wing of \ha line is visible. It is formed by gaseous stream
moving in the direction to the normal star. That suggests the
grazing conjunction in this system. Maximum velocity of the stream is of
650 km/s. Assuming that the stream is moving through the circular Keplerian
orbit around black hole, the mass of the black hole is determined to be
$M_{BH}$=7.1$-$9.5\msun, what confirms the model by Orosz \etal
}
\end{abstract}
V4641 Sgr is a detached stellar system including black hole
with the mass of $8.73 \le M_{BH} \le 11.70$\msun and a normal
B9III type star \cite{Oj01}. In the quiet state, the brightness of the
star is of about 13$^m$.6 $V$. The elements for the inferior
conjunction of the black hole are the following \cite{Gv01}:

$$T_c = 2451764.298 + 2^d.81728 \cdot E.$$

\noindent
The secondary B9 companion shows a strong tidal distortion due to
gravitational influence
of the black hole. The system is known to have a relativistic jet
source \cite{Hr00}. Three historical outbursts
of V4641 Sgr are known, in June 1978, in September 1999, and
in May 2002. In the peak of 1999 outburst, the star reached
brightness of 8$^m$.9 $V$. The high state of 2002 was characterized
by light flickering in the time scale of seconds and minutes
with the amplitude up to 1$^m$.5 and dips \cite{V}. The amplitude of flickering
decreased in the dips.
Here we present the results of optical monitoring and
spectroscopy of V4641 Sgr in the quiet state.

CCD monitoring of V4641 Sgr was done in $V$ and $R$ bands with 60-cm
telescope of SAI Crimean station and with 38-cm telescope of
Crimean Astrophysical Observatory in July-August 2000 and May 2001.
SBIG ST-7 CCD was used.
In the $R$ band, the observations continued about 4 hours each night
and covered well all the phases of orbital period. Fig.~1 shows
light curve of V4641 Sgr, magnitudes are given relative to
comparison star {\bf e} \cite{Gv01}. The accuracy of observations is
equal to 0$^m$.02. The ellipsoidal variations
of high amplitude predominate, and two light minima of unequal
depth are visible. The deepest minimum coinciding with the
phase of superior conjunction of the normal star has amplitude of 0$^m$.36,
the secondary one having only 0$^m$.27. This extreme amplitude suggests
that the system is highly inclined.
The system is considered to be non-eclipsing because no
eclipses are seen in X-ray data. Due to
strong gravitational influence of the  black hole, the surface of the
normal star is elongated mostly in the direction of black hole,
and is heated by central energy source non-uniformly. This may be
the cause of inequality of minima in the light curve. There is no
irradiation effect due to reprocessing of X-rays by normal star
seen in the light curve. There are no features in the primary minimum
which would give evidence of the partial eclipse of the normal star by the
accretion disk, like contacts or light drops.
To ascertain, if surrounding matter exists around
the black hole to absorb light in the spectral lines, we performed
the special spectral observations of V4641 Sgr in the orbital phases
nearby to the inferior conjunction of the black hole.

\begin{figure}[htb]
\centering
\psfig{file=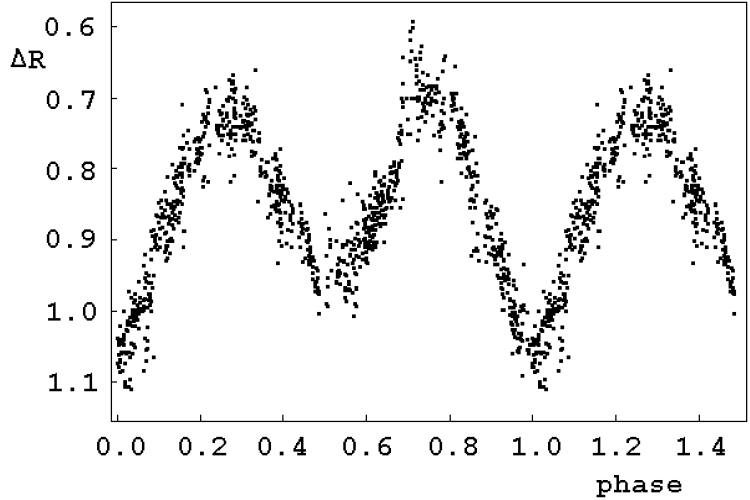,width=10cm}
\caption{ The light curve of V4641 Sgr in R band.}
\label{fig:Figure 1.}
\end{figure}

Spectra of V4641 Sgr were taken with 6-m telescope BTA of Special
Astrophysical Observatory on July 11, 2001 at the orbital phases of
$0^p.9841 - 0^p.0006$. The wavelength ranges were 5800 - 7100\ang\
to check \ha region, and 4630 - 5880\ang\ to overlap \hb and He~II
$\lambda$4686\ang\ regions. The dispersion was 1.25 \ang/pxl, and S/N=200.
Air mass was about 3 atmospheres for this low declination star,
and we observed the nearby A2 type star HD 315568 to control
the Earth atmosphere transparency in the line profiles.
We analised also few spectra of V4641 Sgr taken with 6-m telescope
on August 21, 1996 and on September 17, 1999 in the other orbital
phases.

One hour before the conjunction on July 11, 2001 the depression
with EW = 0.5\ang\ was observed in the red wing of \ha
absorption line (Fig.~2) in two of our spectra. This suggests
the absorption by a gaseous stream moving along the line of site
to the normal star. No such depressions were observed in the \ha line
profile of the control star, no ones in the line profile
of V4641 Sgr on August 1996.

The gaseous stream is assumed to be a part of a rarefied disk
located in the orbital plane (this idea was proposed by S.N. Fabrika).
It is optically thick only in \ha line. Note, that the highest
velocity part of the disk may cover the normal star
at the phase of observation, as follows from a high inclination model
of this binary system. With the absence of X-ray eclipses,
the observation of the stream presumes a "stargrizing"
conjunction and extreme orbit inclination angle of about
i = 70.7\degree. No trace of accretion stream moving from
normal star to black hole is seen in the blue wing of
\ha profile. The observed maximum heliocentric velocity
of the stream of 642 km/s and the center-of-mass velocity of the
system of 110 km/s suggest the calculated velocity $v$ on
the circular Keplerian orbit around the black hole of 650 km/s
at the distance of $R=0.15 - 0.20a$, {\it a} being the distance
between components. The orbital inclination of 70.7\degree,
and the angle between the circular flow direction and the projection
of the line of sight onto orbital plane of about 30\degree
in the region of the star limb covered by the maximum
velocity flow were taken into account in calculations.
Assuming $M_{BH} = v^2R/G$ for mass of central black hole we
have value in the range of $7.1-9.5$\msun what overlaps well
the mass range given by Orosz \etal \cite{Oj01}.

\begin{figure}[htb]
\centering
\psfig{file=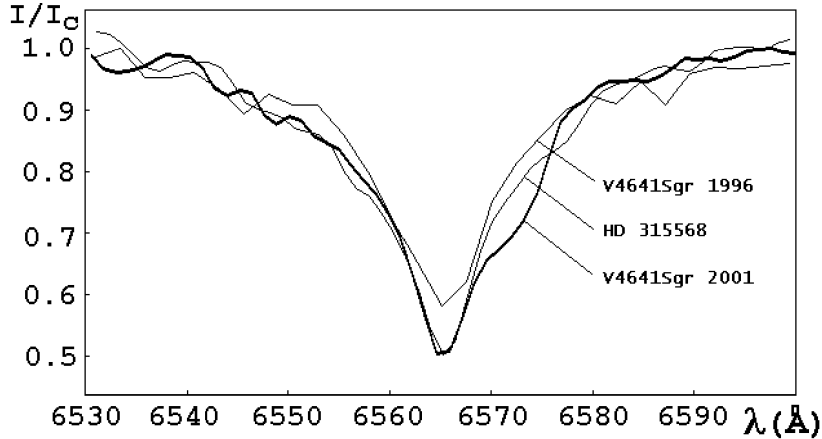,width=11cm}
\caption{
\ha line profile of V4641 Sgr one hour before
the black hole conjunction (solid line). The line profiles of
V4641 Sgr in the phase of $0^p.36$, and of the nearby located
check A2 type star HD 315568 are represented by thin lines.
}
\label{fig:Figure 2.}
\end{figure}

More symmetric line profiles in \hb line were found in the
exact zero phase in the conjunction, but a weak absorption component
may be revealed by Gaussian analysis at the same velocity as that of
\ha. Modelling predicts that
the disk's streams overlapping the light of the secondary star
are visible at large projection angles in zero phase, and the line
profile should be more symmetric.
Model predicts also depression in the blue wing of Balmer lines one
hour after the conjunction. Unfortunately we have no observations in
this orbital phase. There are no  traces
of He~II $\lambda$4686\ang\ line in absorption what suggests,
that the disk is not exposed by X-rays from the
black hole in quiescence.

 The rarefied gaseous disk discovered in this study may be
responsible for supercritical accretion events accompanied
with strong outbursts, when enough matter is accumulated
in it to get the disk instability.

\end{document}